\documentclass[12pt,a4paper]{article}
\usepackage{amsmath}	
\usepackage{amssymb}
\usepackage{graphicx}

\def\xiG{\xi_G}
\def\xiW{\xi_W}
\def\xiB{\xi_B}
\def\alu{\ensuremath{a_t}}
\def\ald{\ensuremath{a_b}}
\def\ale{\ensuremath{a_\tau}}
\def\lam{\ensuremath{\hat\lambda}}

\newcommand{\ala}{\ensuremath {a_1}}
\newcommand{\alb}{\ensuremath {a_2}}
\newcommand{\alc}{\ensuremath {a_s}}

\newcommand{\MS}{{\ensuremath{\overline{\mathrm{MS}}}}}
\def\z#1{{\zeta_{#1}}}
\def\LYukawa{\ensuremath{\mathcal{L}_{\mathrm{Yukawa}}}}
\def\LH{\ensuremath{\mathcal{L}_{\mathrm{H}}}}

\newcommand{\NR}{\ensuremath {N_c}}

\newcommand{\NGen}{\ensuremath {n_G}}

\newcommand{\cA}{\ensuremath {C_A}}
\newcommand{\cR}{\ensuremath {C_F}}

\def \eps {\epsilon}

\begin{document}
\thispagestyle{empty}
\begin{center}
{\Large{\bf
Higgs self-coupling beta-function\\[3mm] in the Standard Model at three loops
}}
\vspace{15mm}

{\sc
A.~V.~Bednyakov${}^1$,
A.~F.~Pikelner${}^1$
and V.~N.~Velizhanin${}^2$}\\[5mm]

${}^1${\it
Joint Institute for Nuclear Research,\\
 141980 Dubna, Russia}

\vspace{5mm}

${}^2${\it
Theoretical Physics Division, Petersburg Nuclear Physics Institute,\\
  Orlova Roscha, Gatchina, 188300 St.~Petersburg, Russia}

\vspace{15mm}

\textbf{Abstract}\\[2mm]
\end{center}

\noindent{
We present the results for three-loop beta-function for the Higgs self-coupling calculated within the unbroken phase of the Standard Model.
We also provide the expression for three-loop beta-function of the Higgs mass parameter,
which is obtained as a by-product of our main calculation.
Our results coincide with that of recent paper arXiv:1303.2890.
In addition, the expression for the Higgs field anomalous dimension is given.
}
\newpage

\setcounter{page}{1}

The Higgs self-interaction coupling being the fundamental parameters of the Standard Model (SM) Lagrangian
	describes the interactions of Higgs field with itself and
        is strongly related to the Higgs mass via electroweak symmetry breaking.
Having in mind the discovery of the Higgs boson~\cite{:2012gk,:2012gu}
	the Higgs self-interaction coupling can be deduced directly
	from the experimental data.
In order to obtain a very precise SM prediction for the running Higgs
self-coupling at some high energy scale, one usually uses value
extracted from Higgs mass measurements around electroweak $M_Z$  scale.
Objects of interest are scales up to the Planck mass, so one inevitably makes use of renormalization group equations (RGE) to connect these scales.
	The SM parameters in such a studies are usually defined in the minimal subtraction (\MS) scheme, in which counter-terms
	and beta-functions have a very simple polynomial structure.
One can use RGE for finding the scale where New Physics should enter the game, e.g.,
	to unify the interactions or stabilize the Higgs potential~\cite{Krasnikov:1978pu,Hung:1979dn,Politzer:1978ic,Bezrukov:2012sa,Degrassi:2012ry,Alekhin:2012py}.

One-
and two-loop results for SM beta functions have been known for quite a long time~\cite{Gross:1973id,Politzer:1973fx,Jones:1974mm,Tarasov:1976ef,Caswell:1974gg,Egorian:1978zx,Jones:1981we,Fischler:1981is,Machacek:1983tz,Machacek:1983fi,Luo:2002ti,Jack:1984vj,Gorishnii:1987ik,Arason:1991ic}
and are summarized in~\cite{Luo:2002ey}.

Not long ago full three-loop beta-functions for gauge couplings~\cite{Mihaila:2012fm,Bednyakov:2012rb} and Yukawa couplings~\cite{Bednyakov:2012en}	were calculated.
The beta-functions for the Higgs self-coupling and top Yukawa coupling were also considered at three loops~\cite{Chetyrkin:2012rz}.
However, in Ref.~\cite{Chetyrkin:2012rz} all the electroweak couplings were neglected together with the Yukawa couplings of
other SM fermions.\footnote{During the preparation of this paper, the authors of Ref.~\cite{Chetyrkin:2012rz} extended their result and incorporate the dependence
  \cite{Chetyrkin:2013wya} on the electroweak gauge couplings and Yukawa couplings.}

In this paper, we provide the full analytical result for the three-loop beta-functions
of the Higgs-self coupling $\lambda$ and the Higgs mass parameter $m^2$.  We
take into account all the interactions of the SM, restricting Yukawa
sector to include only the heaviest fermion generation.

Let us briefly recall our notation.
The full Lagrangian of the unbroken SM which was used in this calculation is given
	in our previous paper~\cite{Bednyakov:2012rb}.
However, we do not keep the full flavor structure of Yukawa interactions but
	use the following simple Lagrangian which describes
        fermion-Higgs interactions and the Higgs field self-interaction
	\begin{eqnarray}
	  \LYukawa &=& - y_t (\bar{Q} \Phi^c) t_R - y_b (\bar{Q}\Phi) b_R - y_\tau (\bar{L} \Phi) \tau_R
	  + \mathrm{h.c.}\,,
	  \label{eq:yukawa_lag}\\[3mm]
          \LH & = & \left( D_\mu \Phi \right)^\dagger  \left( D_\mu \Phi \right) - V_H(\Phi)\,,
	  \label{eq:higgs_lag} \\[1mm]
	  V_H(\Phi) & = &  \lambda \left( \Phi^\dagger \Phi \right)^2 =  \lambda \left( \frac{ h^2 + \chi^2}{2} +  \phi^+ \phi^- \right)^2,
	  \label{eq:higgs_pot}
        \end{eqnarray}
with $Q=(t,b)_L$, and $L=(\nu_\tau, \tau)_L$ being SU(2) doublets of left-handed fermions
of the third generation, $u_R$, $t_R$, and $\tau_R$ are the corresponding right-handed
	counter parts.
The Higgs doublet $\Phi$ with $Y_W = 1$ has the following decomposition in terms of the component fields:
\begin{equation}
	\Phi =
	\left(
	\begin{array}{c}
		\phi^+(x) \\ \frac{1}{\sqrt 2} \left( h + i \chi \right)
		\end{array}
	\right),
	\qquad
	\Phi^c = i\sigma^2 \Phi^\dagger =
	\left(
	\begin{array}{c}
		\frac{1}{\sqrt 2} \left( h - i \chi \right) \\
		-\phi^-
		\end{array}
	\right).
	\label{eq:Phi_def}
\end{equation}
Here a charge-conjugated Higgs doublet is introduced $\Phi^c$ with $Y_W=-1$.
The Higgs self-coupling $\lambda$ entering tree-level Higgs potential~\eqref{eq:higgs_pot}
is of our primary interest.  We do not add a  quadratic (mass) term $m^2 \Phi^\dagger \Phi$ to the potential $V_H$, since the running of the mass parameter $m^2$ can be deduced by considering renormalization of composite operator $\Phi^\dagger \Phi$.
The treatment is essentially the same as in Ref.~\cite{Chetyrkin:2012rz}. Some details can be found below.

For loop calculations it is convenient to define the following quantities:
\begin{eqnarray}
	 a_i  & = &  \left(\frac{5}{3} \frac{g_1^2}{16\pi^2}, \frac{g_2^2}{16\pi^2}, \frac{g_s^2}{16\pi^2},
	 \frac{y_t^2}{16\pi^2}, \frac{y_b^2}{16\pi^2}, \frac{y_\tau^2}{16\pi^2}, \frac{\lambda}{16\pi^2},\xiB, \xiW, \xiG \right),
	 \label{eq:coupl_notations}
\end{eqnarray}
 	where we use the SU(5) normalization of the U(1) gauge coupling $g_1$.
We also stress that the calculation is carried out in a general linear
$R_\xi$ gauge, in which the vector boson propagators has the form
\begin{eqnarray}\label{ksi}
\frac{1}{k^2}\left[g_{\mu\nu}-\hat\xi_Q\frac{k_{\mu}k_{\nu}}{k^2}\right],\qquad\qquad\hat\xi_Q=1-\xi_Q\,.
\end{eqnarray}
A minimal way to test gauge invariance at the end of calculation is to keep at most a
single power of $\hat\xi_Q$, which corresponds to a first order expansion of
the result around the Feynman gauge.

The $\lambda$ beta-function is extracted from the corresponding renormalization constant
which relates the bare coupling to the renormalized one in the~\MS-scheme.
The latter can be found, for example, with the help of the following formulae:
\begin{equation}\label{Zlambda}
	Z_{\lambda} = \frac{Z_{hhhh}}{Z_{h}^2} = \frac{Z_{\chi\chi\chi\chi}}{Z_{\chi}^2}=\frac{Z_{hh\phi^{+}\phi^{-}}}{Z_{h}\sqrt{Z_{\phi^{+}}Z_{\phi^{-}}}}\ ,
\end{equation}
where $Z_{hhhh}$, $Z_{\chi\chi\chi\chi}$, $Z_{hh\phi^{+}\phi^{-}}$ are the renormalization constants for the
        four-point vertices involving four components of the Higgs doublet $\Phi$.

	The renormalization constant $Z_{h}=Z_{\chi}=Z_{\phi^\pm}=Z_{\Phi}$ can be found
	from the corresponding self-energy diagrams.
It turns out that due to the gauge symmetry all Higgs doublet components renormalize in the same way.
Moreover, the same reasoning can be applied to the considered Higgs vertices giving, e.g.,
	$Z_{hhhh} = Z_{\chi\chi\chi\chi} = Z_{hh\phi^+\phi^-}$.

\begin{figure}[t]
   \centering
   \includegraphics[width=\textwidth]{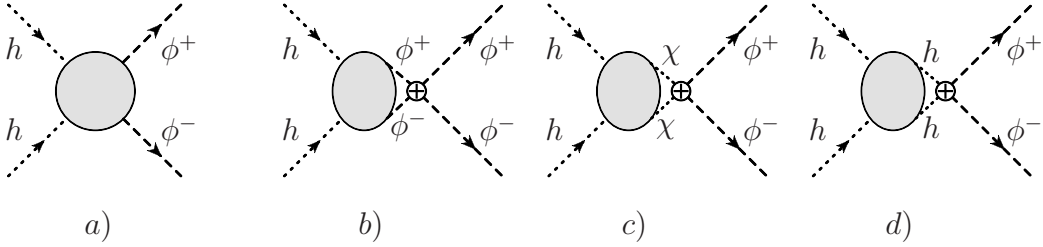}
   \caption{
   For the calculation of the beta-function of the Higgs
self-coupling we evaluate all diagrams included in Fig. $a$.
   For the calculation of the beta-function of the Higgs mass
parameter we should insert the mass operator into self-energy diagrams
for Higgs fields.
   Effectively the mass operator is equivalent to the quartic Higgs
vertices with two external Higgs fields $[\phi^{+}\phi^{-}]$.
  All diagrams corresponding to Fig. $b$ should be multiplied by the factor $1/2$.
   }
   \label{fig:HphHphtomH}
\end{figure}

Renormalization constant for the Higgs mass parameter can be easily
extracted from the calculations of the renormalization of the Higgs
self-coupling. In the most simple way this can be done from our
calculation of $hh\phi^{+}\!\phi^{-}$ vertex. For this purpose we have
labeled all quartic Higgs vertices and have extracted the results,
which contain such vertices with external $\phi^{+}\phi^{-}$ fields.
This trick is illustrated on Fig.~\ref{fig:HphHphtomH}.
Effectively, this trick is equivalent to the insertion of the local operator $O_{2\Phi}=\Phi^\dagger \Phi$, as in Ref.~\cite{Chetyrkin:2012rz}.
The corresponding renormalization constant can be extracted from our results in the following way:
\begin{equation}\label{ZmH}
	Z_{m^2} = \frac{Z_{hh\,[\phi^{+}\phi^{-}]}}{Z_{h}}\ .
\end{equation}
	In order to extract a three-loop contribution to the considered renormalization constants,
	it is sufficient to know the two-loop results for the gauge and Yukawa couplings and the two-loop expression
	for the Higgs self-interaction.

The relation between the bare and renormalized parameters can be written in the following way
\begin{equation}
	a_{k,\mathrm{Bare}}\mu^{-2\rho_k\epsilon} = Z_{a_k} a_k(\mu)=a_k+\sum_{n=1}^\infty c_k^{(n)}\frac{1}{\epsilon^n}\ ,
	\label{eq:bare_to_ren}
\end{equation}
	where $\rho_k=1$ for the gauge and Yukawa constants, $\rho_k=2$ for the scalar quartic coupling constant, and $\rho_k=0$ for the gauge fixing parameters.
	The bare couplings are defined within the dimensionally regularized~\cite{'tHooft:1972fi} theory
	with $D=4-2 \eps$.
The four-dimensional beta-functions, denoted by $\beta_i$, are defined via
\begin{equation}
		\beta_i(a_k) = \frac{d a_i(\mu,\epsilon)}{d \ln \mu^2}\bigg|_{\epsilon=0},\ 		
		\qquad \beta_i = \beta_i^{(1)} + \beta_i^{(2)} + \beta_i^{(3)} + \ldots
		\label{eq:beta_def}
\end{equation}
	with $\beta^{(l)}_i$ being the $l$-loop contribution to the beta-function for $a_i$.
	The expression for $\beta_i$  can be extracted from the corresponding renormalization constants~\eqref{eq:bare_to_ren} with the help of
\begin{equation}
 	\beta_i =  \sum_{l}\rho_l a_l\frac{\partial c_i^{(1)}}{\partial a_l}-\rho_i c_i^{(1)}\,.
	 	\label{eq:beta_calc_simple}
	\end{equation}
	Here, again, $a_i$ stands for both the gauge couplings and the gauge-fixing.

As in our previous work~\cite{Velizhanin:2008jd,Velizhanin:2010vw,Velizhanin:2011es}
all calculations were performed with FORM~\cite{Vermaseren:2000nd}, using FORM package
COLOR~\cite{vanRitbergen:1998pn} for evaluation of the color traces.
Feynman integrals are evaluated by the method from
Refs.~\cite{Misiak:1994zw,Chetyrkin:1997fm} and our own implementation
of the Laporta's algorithm~\cite{Laporta:2001dd} in the form of the MATHEMATICA package
BAMBA with the master integrals from Ref.~\cite{Czakon:2004bu}.

According to the prescription of Refs.~\cite{Misiak:1994zw,Chetyrkin:1997fm}, we introduce an auxiliary mass parameter $M$
in all propagator denominators and perform an expansion in external momenta.
Due to this, we only have to consider vacuum integrals with one mass scale. The subtlety of the method is related to the fact that
one needs to introduce mass counter-terms for all the boson fields of the model, i.e., gauge and Higgs fields.
Moreover, we have to consider diagrams with counter-term insertions for all the vertices of the model.

\begin{table}[t]
\centering
\vskip 2mm
\begin{tabular}{|c|c|c|c|}
 \hline
                                     & 1-loop & 2-loop  & 3-loop \\
   \hline
$hhhh$                                & 246    & 40\,905 & 8\,659\,436 \\
 \hline
$\chi\chi\chi\chi$                    & 246    & 40\,905 & 8\,659\,436 \\
 \hline
$hh\phi^{+}\phi^{-}$                  & 146    & 29\,289 & 6\,741\,584 \\
 \hline
$\chi\chi\phi^{+}\phi^{-}$            & 146    & 29\,289 & 6\,741\,584 \\
 \hline
$\phi^{+}\phi^{-}\phi^{+}\phi^{-}$    & 193    & 35\,211 & 7\,597\,252 \\
 \hline
$hh\chi\chi$                          & 168    & 32\,469 & 7\,378\,694 \\
 \hline
 \hline
$hh$                                  & 9      & 419     & 41\,369  \\
 \hline
$\chi\chi$                            & 9      & 419     & 41\,369  \\
 \hline
$\phi^{+}\phi^{-}$                    & 8      & 394     & 39\,122  \\
 \hline
 \hline
$hh\;[\phi^{+}\phi^{-}]$              & 9      & 900     & 140\,979 \\
 \hline
\end{tabular}
\caption{The number of diagrams for calculations up to three-loop order.}
\label{numbdiagr}
\end{table}

For dealing with a huge number of diagrams (see Table~\ref{numbdiagr})
instead of FeynArts~\cite{Hahn:2000kx} package, exploited in our recent
studies\cite{Bednyakov:2012rb,Bednyakov:2012en}, we use a program DIANA~\cite{Tentyukov:1999is},
which calls QGRAF~\cite{Nogueira:1991ex} to generate all
diagrams. By means of a prepared script we map topologies generated by DIANA
on previously defined auxiliary topology for IBP identities.
The model file for the unbroken SM, used previously with FeynArts,
was converted to the DIANA model format, which allow us to introduce vertex counter-terms in a convenient way.
Most of these counter-terms were generated from the known two-loop renormalization constants for SM parameters and fields.
We have calculated $\chi\chi\chi\chi$ and $hh\phi^{+}\phi^{-}$ vertices and the obtained results are the same. The last vertex was used for the calculation of the $\beta$-function for the mass parameter. Both calculations were performed in the linear gauge (see Eq.~(\ref{ksi})) and we keep only the first power of $\hat\xi_Q$.
These results can be found online as ancillary files of the \texttt{arXiv} version of the paper.
Since we are considering three-loop Green function with scalar external legs at zero external momenta it is easy to convince oneself (see, e.g., reasoning
given in Refs.~\cite{Mihaila:2012pz,Chetyrkin:2012rz,Bednyakov:2012rb,Chetyrkin:2013wya})
that the naive anticommuting prescription for the $\gamma_5$ matrix is sufficient for our current study.

As a result of our calculation we obtain the expressions
for the three-loop Higgs self-coupling \eqref{eq:betalam_3} and mass parameter \eqref{eq:betam2_3} beta-functions ($\lam\equiv a_\lambda$):
{\allowdisplaybreaks
\begin{align}
	\beta^{(1)}_{\lambda} & =
12 \lam^2
-\frac{9 \ala \lam}{10}
-\frac{9 \alb \lam}{2}
+\frac{27 \ala^2}{400}
+\frac{9 \ala \alb}{40}
+\frac{9 \alb^2}{16}
\nonumber\\
&
-3 \alu^2
-3 \ald^2
-\ale^2
+6 \alu \lam
+6 \ald \lam
+2 \ale \lam
\label{eq:betalam_1}\,,
\\
	\beta^{(2)}_{\lambda} & =
\alb^3 \Big(\frac{497}{32}-2 \NGen\Big)
+\ala^3 \Big(-\frac{6 \NGen}{25}-\frac{531}{4000}\Big)
+\alb^2 \lam \Big(5 \NGen-\frac{313}{16}\Big)
\nonumber\\
&
+\ala^2 \lam \Big(\NGen+\frac{687}{400}\Big)
+\ala^2 \alb \Big(-\frac{2 \NGen}{5}-\frac{717}{800}\Big)
+\ala \alb^2 \Big(-\frac{2 \NGen}{5}-\frac{97}{160}\Big)
\nonumber\\
&
-\frac{171 \ala^2 \alu}{200}
+\frac{9 \ala^2 \ald}{40}
-\frac{9 \ala^2 \ale}{8}
-\frac{9 \alb^2 \alu}{8}
-\frac{9 \alb^2 \ald}{8}
-\frac{3 \alb^2 \ale}{8}
+\frac{27 \ala \alb \ald}{20}
\nonumber\\
&
+\frac{33 \ala \alb \ale}{20}
+\frac{63 \ala \alb \alu}{20}
-\frac{4 \ala \alu^2}{5}
+\frac{2 \ala \ald^2}{5}
-\frac{6 \ala \ale^2}{5}
+15 \alu^3
+15 \ald^3
+5 \ale^3
\nonumber\\
&
-16 \alc \ald^2
-16 \alc \alu^2
+\frac{117 \ala \alb \lam}{40}
+\frac{17 \ala \alu \lam}{4}
+\frac{5 \ala \ald \lam}{4}
+\frac{15 \ala \ale \lam}{4}
\nonumber\\
&
-3 \ald^2 \alu
-3 \ald \alu^2
-\frac{3 \ald^2 \lam}{2}
-\frac{3 \alu^2 \lam}{2}
-\frac{\ale^2 \lam}{2}
+\frac{45 \alb \alu \lam}{4}
+\frac{45 \alb \ald \lam}{4}
\nonumber\\
&
+\frac{15 \alb \ale \lam}{4}
+40 \alc \ald \lam
+40 \alc \alu \lam
-21 \ald \alu \lam
+\frac{54 \ala \lam^2}{5}
\nonumber\\
&
+54 \alb \lam^2
-72 \alu \lam^2
-72 \ald \lam^2
-24 \ale \lam^2
-156 \lam^3
\label{eq:betalam_2}\,,
\\
	\beta^{(3)}_{\lambda} & =
\ala^3 \alb \Big(-\frac{2 \NGen^2}{9}+\NGen \Big(\frac{183 \z3}{125}-\frac{18001}{12000}\Big)+\frac{81 \z3}{160}-\frac{29779}{32000}\Big)
-\frac{27}{5} \ala \ale \alu \lam
\nonumber\\
&
+\ala^2 \alb^2 \Big(-\frac{2 \NGen^2}{9}+\NGen \Big(\frac{63 \z3}{25}+\frac{149}{1800}\Big)+\frac{7857 \z3}{1600}-\frac{64693}{9600}\Big)
-\frac{27}{5} \ala \ald \ale \lam
\nonumber\\
&
+\ala^4 \Big(-\frac{\NGen^2}{5}+\NGen \Big(\frac{171 \z3}{125}-\frac{12441}{8000}\Big)+\frac{8019 \z3}{80000}-\frac{12321}{128000}\Big)
 +\frac{123}{200} \ala^2 \ald \ale
\nonumber\\
&
+\ala^3 \lam \Big(\frac{14 \NGen^2}{9}-\frac{114 \NGen \z3}{25}+\frac{1199 \NGen}{150}+\frac{243 \z3}{1000}+\frac{12679}{2000}\Big)
+\frac{2103}{200} \ala^2 \ale \alu
\nonumber\\
&
+\alb^4 \Big(-\frac{5 \NGen^2}{3}+\NGen \Big(-45 \z3-\frac{14749}{192}\Big)-\frac{2781 \z3}{128}+\frac{982291}{3072}\Big)
+21 \ald \ale \alu \lam
\nonumber\\
&
+\alb^3 \lam \Big(\frac{70 \NGen^2}{9}+90 \NGen \z3+\frac{3515 \NGen}{36}+\frac{2259 \z3}{8}-\frac{46489}{288}\Big)
+\frac{87}{20} \ala \alb \ale \alu
\nonumber\\
&
+\ala^2 \alb \lam \Big(-\frac{54 \NGen \z3}{25}+\frac{171 \NGen}{20}-\frac{27 \z3}{200}+\frac{8811}{200}\Big)
+\ale^4 \Big(-12 \z3-\frac{143}{8}\Big)
\nonumber\\
&
+\ala \alb^2 \lam \Big(-\frac{18 \NGen \z3}{5}+\frac{99 \NGen}{10}-\frac{747 \z3}{40}+\frac{13659}{160}\Big)
+\lam^4 (2016 \z3+3588)
\nonumber\\
&
+\ala \alb^3 \Big(-\frac{2 \NGen^2}{9}-\frac{8341 \NGen}{1440}-\frac{243 \z3}{32}-\frac{54053}{5760}\Big)
+\frac{9}{8} \alb^2 \ale \alu
+\frac{9}{8} \alb^2 \ald \ale
\nonumber\\
&
+\ala^2 \alu \lam \Big(-\frac{127 \NGen}{20}-\frac{1347 \z3}{50}-\frac{112447}{4800}\Big)
+\frac{45}{8} \ald^2 \ale \alu
+\frac{45}{8} \ald \ale \alu^2
\nonumber\\
&
+\ala^2 \ale \lam \Big(-\frac{117 \NGen}{20}-\frac{1107 \z3}{50}-\frac{16047}{1600}\Big)
+\frac{477}{32} \alb \ald^2 \alu
+\frac{477}{32} \alb \ald \alu^2
\nonumber\\
&
+\ala^2 \ald \lam \Big(-\frac{31 \NGen}{20}-\frac{141 \z3}{50}-\frac{127303}{4800}\Big)
+\alc \alu^2 \lam (895-1296 \z3)
\nonumber\\
&
+\ala^2 \alu^2 \Big(-\frac{23 \NGen}{20}+\frac{2957 \z3}{400}+\frac{100913}{9600}\Big)
+\alc \ald^2 \lam (895-1296 \z3)
\nonumber\\
&
+\ala^2 \alb \ale \Big(-\frac{3 \NGen}{10}-\frac{27 \z3}{10}+\frac{59913}{6400}\Big)
+\alc \ald \lam^2 (1152 \z3-1224)
\nonumber\\
&
+\ala^2 \ald^2 \Big(-\frac{83 \NGen}{20}-\frac{407 \z3}{80}+\frac{15137}{9600}\Big)
+\alc \alu \lam^2 (1152 \z3-1224)
\nonumber\\
&
+\alb^2 \alu^2 \Big(-\frac{39 \NGen}{4}-\frac{819 \z3}{16}+\frac{13653}{128}\Big)
+\ald \alu \lam^2 (117-864 \z3)
\nonumber\\
&
+\alb^2 \alu \lam \Big(-\frac{63 \NGen}{4}-\frac{351 \z3}{2}-\frac{3933}{64}\Big)
+\alb \alc \ald \alu (96 \z3-8)
\nonumber\\
&
+\alb^2 \ale \lam \Big(-\frac{21 \NGen}{4}-\frac{117 \z3}{2}-\frac{1311}{64}\Big)
+\alc \alu^3 (240 \z3-38)
\nonumber\\
&
+\alb^2 \ald \lam \Big(-\frac{63 \NGen}{4}-\frac{351 \z3}{2}-\frac{3933}{64}\Big)
+\alc \ald^3 (240 \z3-38)
\nonumber\\
&
+\alb^2 \ald^2 \Big(-\frac{39 \NGen}{4}-\frac{819 \z3}{16}+\frac{13653}{128}\Big)
-72 \ale^2 \alu^2
-72 \ald^2 \ale^2
+72 \ald^2 \alu^2 \z3
\nonumber\\
&
+\ala^3 \ale \Big(\frac{99 \NGen}{50}-\frac{81 \z3}{100}+\frac{106083}{32000}\Big)
+\alc \ald \alu \lam (82-96 \z3)
+291 \ale \lam^3
\nonumber\\
&
+\ala^2 \alb \alu \Big(\frac{3 \NGen}{10}-\frac{27 \z3}{25}+\frac{70563}{6400}\Big)
+\ala^3 \alu \Big(\frac{129 \NGen}{50}-\frac{27 \z3}{50}+\frac{128829}{32000}\Big)
\nonumber\\
&
+\ala^2 \alb \ald \Big(\frac{3 \NGen}{2}+\frac{81 \z3}{50}+\frac{39627}{6400}\Big)
+\ala^3 \ald \Big(\frac{57 \NGen}{50}+\frac{27 \z3}{100}+\frac{36129}{32000}\Big)
\nonumber\\
&
+\ala \alb^2 \alu \Big(\frac{3 \NGen}{10}+\frac{81 \z3}{20}+\frac{9309}{1280}\Big)
+\ala^2 \ale^2 \Big(\frac{39 \NGen}{20}+\frac{135 \z3}{16}+\frac{51273}{3200}\Big)
\nonumber\\
&
+\ala \alb^2 \ale \Big(-\frac{3 \NGen}{10}-\frac{9 \z3}{10}+\frac{5499}{1280}\Big)
+\ala \alb^2 \ald \Big(\frac{3 \NGen}{2}+\frac{27 \z3}{5}+\frac{12537}{1280}\Big)
\nonumber\\
&
+\alb^2 \ale^2 \Big(-\frac{13 \NGen}{4}-\frac{273 \z3}{16}+\frac{4503}{128}\Big)
+\ala \alb \lam^2 \Big(-\frac{486 \z3}{5}-\frac{999}{5}\Big)
\nonumber\\
&
+\ala^2 \lam^2 \Big(-\frac{141 \NGen}{5}-\frac{729 \z3}{25}-\frac{1647}{25}\Big)
+\alb^3 \alu \Big(\frac{27 \NGen}{2}+\frac{297 \z3}{4}-\frac{17217}{256}\Big)
\nonumber\\
&
+\alb^3 \ald \Big(\frac{27 \NGen}{2}+\frac{297 \z3}{4}-\frac{17217}{256}\Big)
+\alb^3 \ale \Big(\frac{9 \NGen}{2}+\frac{99 \z3}{4}-\frac{5739}{256}\Big)
\nonumber\\
&
+\ala^2 \alc \lam \Big(\frac{99 \NGen}{10}-\frac{264 \NGen \z3}{25}\Big)
+\alb^2 \ald \alu \Big(-12 \NGen+\frac{117 \z3}{2}-\frac{351}{64}\Big)
\nonumber\\
&
+\ala^2 \alb \alc \NGen \Big(\frac{66 \z3}{25}-\frac{561}{200}\Big)
+\ala^3 \alc \NGen \Big(\frac{198 \z3}{125}-\frac{1683}{1000}\Big)
+873 \alu \lam^3
\nonumber\\
&
+\ala \alb \alu \lam \Big(\frac{531 \z3}{5}-\frac{19527}{160}\Big)
+\ala \alb \ale \lam \Big(\frac{378 \z3}{5}-\frac{11313}{160}\Big)
+873 \ald \lam^3
\nonumber\\
&
+\ala \alb^2 \alc \NGen \Big(\frac{18 \z3}{5}-\frac{153}{40}\Big)
+\ala \alb \alu^2 \Big(-\frac{2229 \z3}{40}-\frac{1079}{320}\Big)
+192 \alc^2 \ald \alu
\nonumber\\
&
+\ala \alb \ald \alu \Big(\frac{93 \z3}{10}+\frac{1001}{160}\Big)
+\ala \alc \alu \lam \Big(\frac{408 \z3}{5}-\frac{2419}{30}\Big)
-\frac{3}{4} \ala \alb \ald \ale
\nonumber\\
&
+\ala \alb \ald^2 \Big(-\frac{933 \z3}{40}-\frac{3239}{320}\Big)
+\ala \alb \ald \lam \Big(\frac{36 \z3}{5}-\frac{9027}{160}\Big)
+12 \ald \ale^2 \alu
\nonumber\\
&
+\ala^2 \ald \alu \Big(-\frac{9 \z3}{25}-\frac{6381}{1600}\Big)
+\ala^2 \alc \alu \Big(\frac{1761}{200}-\frac{162 \z3}{25}\Big)
-216 \ale \alu \lam^2
\nonumber\\
&
+\ala^2 \alc \ald \Big(\frac{2049}{200}-\frac{162 \z3}{25}\Big)
+\ala \alb \alc \alu \Big(\frac{747}{20}-\frac{108 \z3}{5}\Big)
-216 \ald \ale \lam^2
\nonumber\\
&
+\ala \alb \alc \ald \Big(\frac{699}{20}-\frac{108 \z3}{5}\Big)
+\ala \ald \alu \lam \Big(-\frac{6 \z3}{5}-\frac{929}{20}\Big)
+240 \ald^2 \ale \lam
\nonumber\\
&
+\ala \ale^2 \lam \Big(\frac{1521}{40}-\frac{351 \z3}{5}\Big)
+\ala \ale \lam^2 \Big(\frac{288 \z3}{5}-\frac{1623}{20}\Big)
+240 \ald \ale^2 \lam
\nonumber\\
&
+\ala \ald^2 \lam \Big(\frac{747 \z3}{5}-\frac{5737}{40}\Big)
+\ala \ald^2 \alu \Big(\frac{78 \z3}{5}-\frac{2299}{160}\Big)
+240 \ale^2 \alu \lam
\nonumber\\
&
+\ala \ald \lam^2 \Big(\frac{1251}{20}-\frac{576 \z3}{5}\Big)
+\ala \ald \alu^2 \Big(\frac{1337}{160}-\frac{84 \z3}{5}\Big)
+240 \ale \alu^2 \lam
\nonumber\\
&
+\ala \alb \ale^2 \Big(-\frac{1143 \z3}{40}-\frac{9}{64}\Big)
+\alb^2 \alc \lam \Big(\frac{135 \NGen}{2}-72 \NGen \z3\Big)
-27 \alb \ale \alu \lam
\nonumber\\
&
+\ala \alu \lam^2 \Big(-\frac{144 \z3}{5}-\frac{117}{4}\Big)
+\ala \alc \ald^2 \Big(\frac{136 \z3}{5}-\frac{641}{30}\Big)
-27 \alb \ald \ale \lam
\nonumber\\
&
+\alc^2 \alu \lam \Big(-64 \NGen-48 \z3+\frac{1820}{3}\Big)
+\alc^2 \ald \lam \Big(-64 \NGen-48 \z3+\frac{1820}{3}\Big)
\nonumber\\
&
+\alb^2 \lam^2 \Big(-141 \NGen-513 \z3+\frac{1995}{8}\Big)
+\ala \alu^3 \Big(\frac{51 \z3}{5}+\frac{3467}{160}\Big)
-\frac{297 \ald \ale^3}{8}
\nonumber\\
&
+\alc^2 \ald^2 \Big(40 \NGen+32 \z3-\frac{626}{3}\Big)
+\alc^2 \alu^2 \Big(40 \NGen+32 \z3-\frac{626}{3}\Big)
-\frac{297 \ald^3 \ale}{8}
\nonumber\\
&
+\alb \alc \alu \lam \Big(216 \z3-\frac{489}{2}\Big)
+\alb \alc \ald \lam \Big(216 \z3-\frac{489}{2}\Big)
-\frac{297 \ale^3 \alu}{8}
\nonumber\\
&
+\ala \alc \ald \lam \Big(24 \z3-\frac{991}{30}\Big)
+\alb \ald \alu \lam \Big(54 \z3-\frac{531}{4}\Big)
-\frac{297 \ale \alu^3}{8}
\nonumber\\
&
+\ald^2 \alu \lam \Big(144 \z3+\frac{6399}{8}\Big)
+\ald \alu^2 \lam \Big(144 \z3+\frac{6399}{8}\Big)
-\alc \ald^2 \alu (48 \z3+2)
\nonumber\\
&
+\alb \alu^2 \lam \Big(513 \z3-\frac{4977}{8}\Big)
+\alb \ale^2 \lam \Big(171 \z3-\frac{1587}{8}\Big)
-\alc \ald \alu^2 (48 \z3+2)
\nonumber\\
&
+\alb \ald^2 \lam \Big(513 \z3-\frac{4977}{8}\Big)
+\alb^3 \alc \NGen \Big(18 \z3-\frac{153}{8}\Big)
+\alb \lam^3 (72 \z3-474)
\nonumber\\
&
+\ala \lam^3 \Big(\frac{72 \z3}{5}-\frac{474}{5}\Big)
+\ala \ale^3 \Big(\frac{99 \z3}{5}+\frac{81}{32}\Big)
+\ala \alu^2 \lam \Big(\frac{171 \z3}{5}-\frac{497}{8}\Big)
\nonumber\\
&
+\alb \alu \lam^2 \Big(\frac{639}{4}-432 \z3\Big)
+\alb \ale \lam^2 \Big(\frac{213}{4}-144 \z3\Big)
+\alb \ald \lam^2 \Big(\frac{639}{4}-432 \z3\Big)
\nonumber\\
&
+\alb^2 \alc \alu \Big(\frac{651}{8}-54 \z3\Big)
+\alb^2 \alc \ald \Big(\frac{651}{8}-54 \z3\Big)
+\alu^2 \lam^2 \Big(756 \z3+\frac{1719}{2}\Big)
\nonumber\\
&
+\ald^2 \lam^2 \Big(756 \z3+\frac{1719}{2}\Big)
+\alb \alc \alu^2 \Big(24 \z3-\frac{31}{2}\Big)
+\alb \alc \ald^2 \Big(24 \z3-\frac{31}{2}\Big)
\nonumber\\
&
+\ale^2 \lam^2 \Big(252 \z3+\frac{717}{2}\Big)
+\ala \ald^3 \Big(\frac{5111}{160}-15 \z3\Big)
+\ale^3 \lam \Big(-66 \z3-\frac{1241}{8}\Big)
\nonumber\\
&
+\alb \alu^3 \Big(\frac{3411}{32}-27 \z3\Big)
+\alb \ald^3 \Big(\frac{3411}{32}-27 \z3\Big)
+\alu^3 \lam \Big(\frac{117}{8}-198 \z3\Big)
\nonumber\\
&
+\ald^3 \lam \Big(\frac{117}{8}-198 \z3\Big)
+\ald^3 \alu \Big(\!\!-36 \z3-\frac{717}{8}\Big)
+\ald \alu^3 \Big(\!\!-36 \z3-\frac{717}{8}\Big)
\nonumber\\
&
+\alb \ale^3 \Big(\frac{1137}{32}-9 \z3\Big)
+\alu^4 \Big(\!\!-36 \z3-\frac{1599}{8}\Big)
+\ald^4 \Big(\!\!-36 \z3-\frac{1599}{8}\Big),
\!\!\label{eq:betalam_3}
\end{align}

\begin{align}
\frac{\beta^{(1)}_{m^2}}{m^2} & =
-\frac{9 \ala}{20}
-\frac{9 \alb}{4}
+3 \alu
+3 \ald
+\ale
+6 \lam
\label{eq:betam2_1}\,,
\\
	\frac{\beta^{(2)}_{m^2}}{m^2} & =
\ala^2 \Big(\frac{\NGen}{2}+\frac{471}{800}\Big)
+\alb^2 \Big(\frac{5 \NGen}{2}-\frac{385}{32}\Big)
+\frac{36 \ala \lam}{5}
+36 \alb \lam
-30 \lam^2
-36 \alu \lam
\nonumber\\
&
-36 \ald \lam
-12 \ale \lam
+\frac{9 \ala \alb}{16}
+\frac{17 \ala \alu}{8}
+\frac{5 \ala \ald}{8}
+\frac{15 \ala \ale}{8}
+\frac{45 \alb \alu}{8}
-\frac{9 \ale^2}{4}
\nonumber\\
&
+\frac{45 \alb \ald}{8}
+\frac{15 \alb \ale}{8}
+20 \alc \alu
+20 \alc \ald
-\frac{27 \ald^2}{4}
-\frac{21 \ald \alu}{2}
-\frac{27 \alu^2}{4}\ ,
\label{eq:betam2_2}
\\
	\frac{\beta^{(3)}_{m^2}}{m^2} & =
1026 \lam^3
+72 \ale^2 \alu
+72 \ale \alu^2
-108 \ald \ale \lam
-108 \ale \alu \lam
\nonumber\\
&
+\frac{297 \alu \lam^2}{2}
+\frac{297 \ald \lam^2}{2}
+\frac{99 \ale \lam^2}{2}
-\alb \lam^2 (108 \z3+63)
-\frac{27}{2} \alb \ale \alu
\nonumber\\
&
+\alc \ald^2 \Big(\frac{447}{2}-360 \z3\Big)
+\alc \alu^2 \Big(\frac{447}{2}-360 \z3\Big)
+\alc \ald \alu (41-48 \z3)
\nonumber\\
&
+\alb \ald \lam \Big(\frac{567}{8}-324 \z3\Big)
+\alb \alu \lam \Big(\frac{567}{8}-324 \z3\Big)
+\alb \ald \alu \Big(-27 \z3-\frac{243}{8}\Big)
\nonumber\\
&
+\alb^2 \lam \Big(-\frac{153 \NGen}{2}-162 \z3+\frac{11511}{32}\Big)
-\ald \alu \lam \Big(216 \z3+\frac{315}{2}\Big)
\nonumber\\
&
+\alb \ale \lam \Big(\frac{189}{8}-108 \z3\Big)
+\ala \ald \lam \Big(\frac{1179}{40}-\frac{396 \z3}{5}\Big)
+\ala \ale^2 \Big(\frac{873}{80}-\frac{108 \z3}{5}\Big)
\nonumber\\
&
+\alb^2 \ald \Big(-\frac{63 \NGen}{8}-\frac{243 \z3}{4}-\frac{765}{128}\Big)
+\alb^2 \alu \Big(-\frac{63 \NGen}{8}-\frac{243 \z3}{4}-\frac{765}{128}\Big)
\nonumber\\
&
+\ala \alu \lam \Big(-36 \z3-\frac{657}{40}\Big)
+\alb^2 \alc \NGen \Big(\frac{135}{4}-36 \z3\Big)
+\frac{21 \ald \ale \alu}{2}
\nonumber\\
&
+\alc^2 \ald \Big(-32 \NGen-24 \z3+\frac{910}{3}\Big)
+\alc^2 \alu \Big(-32 \NGen-24 \z3+\frac{910}{3}\Big)
\nonumber\\
&
+\ala \lam^2 \Big(-\frac{108 \z3}{5}-\frac{63}{5}\Big)
+\alb^2 \ale \Big(-\frac{21 \NGen}{8}-\frac{81 \z3}{4}-\frac{255}{128}\Big)
\nonumber\\
&
+\ala^2 \lam \Big(-\frac{153 \NGen}{10}-\frac{162 \z3}{25}-\frac{9693}{800}\Big)
+\ala^2 \alc \NGen \Big(\frac{99}{20}-\frac{132 \z3}{25}\Big)
\nonumber\\
&
+\ala^2 \alb \Big(\NGen \Big(\frac{99}{40}-\frac{27 \z3}{25}\Big)-\frac{1863 \z3}{400}+\frac{9477}{800}\Big)
+\ale^3 \Big(15 \z3-\frac{233}{16}\Big)
\nonumber\\
&
+\ala^2 \alu \Big(-\frac{127 \NGen}{40}-\frac{447 \z3}{100}-\frac{123103}{9600}\Big)
+\ala \alc \ald \Big(12 \z3-\frac{991}{60}\Big)
\nonumber\\
&
+\ala^2 \ale \Big(-\frac{117 \NGen}{40}-\frac{27 \z3}{20}-\frac{32463}{3200}\Big)
+\ala \alu^2 \Big(\frac{36 \z3}{5}-\frac{1293}{80}\Big)
\nonumber\\
&
+\ala^2 \ald \Big(-\frac{31 \NGen}{40}-\frac{21 \z3}{20}-\frac{79207}{9600}\Big)
+\ala \ald \alu \Big(-\frac{3 \z3}{5}-\frac{929}{40}\Big)
\nonumber\\
&
+\ala \alb \lam \Big(\frac{108 \z3}{5}-\frac{5103}{80}\Big)
+\ala \ale \lam \Big(\frac{108 \z3}{5}-\frac{1647}{40}\Big)
+\ald^3 \Big(45 \z3+\frac{1605}{16}\Big)
\nonumber\\
&
+\ala \alb \ale \Big(27 \z3-\frac{6993}{320}\Big)
+\ala \alb \alu \Big(\frac{351 \z3}{10}-\frac{9831}{320}\Big)
+\alu^3 \Big(45 \z3+\frac{1605}{16}\Big)
\nonumber\\
&
+\ald^2 \alu \Big(36 \z3+\frac{4047}{16}\Big)
+\ald \alu^2 \Big(36 \z3+\frac{4047}{16}\Big)
+\ala \alc \alu \Big(\frac{204 \z3}{5}-\frac{2419}{60}\Big)
\nonumber\\
&
+\ala \ald^2 \Big(\frac{216 \z3}{5}-\frac{3201}{80}\Big)
+\alb \ale^2 \Big(54 \z3-\frac{987}{16}\Big)
+\ale^2 \lam \Big(72 \z3+\frac{261}{4}\Big)
\nonumber\\
&
+\alb \alc \ald \Big(108 \z3-\frac{489}{4}\Big)
+\alb \alc \alu \Big(108 \z3-\frac{489}{4}\Big)
+72 \ald^2 \ale
+72 \ald \ale^2
\nonumber\\
&
+\alb \ald^2 \Big(162 \z3-\frac{3177}{16}\Big)
+\alb \alu^2 \Big(162 \z3-\frac{3177}{16}\Big)
 -\frac{519}{64} \ala \alb \ald
\nonumber\\
&
+\ald^2 \lam \Big(216 \z3+\frac{351}{4}\Big)
+\alu^2 \lam \Big(216 \z3+\frac{351}{4}\Big)
-\frac{27}{10} \ala \ald \ale
-\frac{27}{2} \alb \ald \ale
\nonumber\\
&
+\alc \ald \lam (576 \z3-612)
+\alc \alu \lam (576 \z3-612)
\nonumber\\
&
+\alb^3 \Big(\frac{35 \NGen^2}{9}+\NGen \Big(45 \z3+\frac{2867}{72}\Big)+\frac{711 \z3}{16}-\frac{39415}{576}\Big)
\label{eq:betam2_3}
\end{align}
}
with $\NGen$ corresponding to the number of SM generations and $\z3 = \zeta(3)$. To save space we substitute all
the color invariants by the corresponding values ($\cR = 4/3, \NR = 3, \cA=3$).
These results with all color invariants can be found online as ancillary files of the \texttt{arXiv} version of the paper.

In addition, we present the expression for the Higgs field anomalous dimension in the Landau gauge, which is usually adopted in the effective potential calculations \cite{Nielsen:1975fs,Fukuda:1975di}\footnote{The result in a general linear $R_\xi$ gauge can be found online as ancillary file of the \texttt{arXiv} version of the paper.}:
{\allowdisplaybreaks
\begin{align}
\gamma^{(1)}_{\Phi} & =
-\frac{9 \ala}{20}
-\frac{9 \alb}{4}
+3 \alu
+3 \ald
+\ale\,,
\label{eq:gamH_1}
\\
\gamma^{(2)}_{\Phi} & =
\ala^2 \Big(\frac{\NGen}{2}+\frac{93}{800}\Big)
+\alb^2 \Big(\frac{5 \NGen}{2}-\frac{511}{32}\Big)
+\frac{27 \ala \alb}{80}
+\frac{17 \ala \alu}{8}
+\frac{5 \ala \ald}{8}
\nonumber\\
&
+\frac{15 \ala \ale}{8}
+\frac{45 \alb \alu}{8}
+\frac{45 \alb \ald}{8}
+\frac{15 \alb \ale}{8}
+20 \alc \alu
+20 \alc \ald
+6 \lam^2
\nonumber\\
&
+\frac{3 \ald \alu}{2}
-\frac{27 \alu^2}{4}
-\frac{27 \ald^2}{4}
-\frac{9 \ale^2}{4}\ ,
\label{eq:gamH_2}
\\
\gamma^{(3)}_{\Phi} & =
\ala^3 \Big(\frac{7 \NGen^2}{9}+\NGen \Big(\frac{158}{75}-\frac{57 \z3}{25}\Big)-\frac{81 \z3}{2000}+\frac{413}{2000}\Big)
+\ala^2 \alc \NGen \Big(\frac{99}{20}-\frac{132 \z3}{25}\Big)
\nonumber\\
&
+\alb^3 \Big(\frac{35 \NGen^2}{9}+\NGen \Big(45 \z3+\frac{2381}{72}\Big)-\frac{207 \z3}{16}-\frac{70519}{576}\Big)
+\alc \ald \alu (57-48 \z3)
\nonumber\\
&
+\ala^2 \alb \Big(\NGen \Big(\frac{9}{40}-\frac{27 \z3}{25}\Big)+\frac{81 \z3}{400}+\frac{837}{800}\Big)
+\ala \alb \alu \Big(\frac{81 \z3}{10}+\frac{1113}{320}\Big)
\nonumber\\
&
+\ala \alb^2 \Big(\NGen \Big(\frac{9}{10}-\frac{9 \z3}{5}\Big)-\frac{27 \z3}{80}+\frac{153}{64}\Big)
+\ala \alc \alu \Big(\frac{204 \z3}{5}-\frac{2419}{60}\Big)
\nonumber\\
&
+\ala^2 \alu \Big(-\frac{127 \NGen}{40}+\frac{3 \z3}{100}-\frac{52831}{9600}\Big)
+\ala^2 \ale \Big(-\frac{117 \NGen}{40}+\frac{351 \z3}{100}-\frac{25551}{3200}\Big)
\nonumber\\
&
+\ala^2 \ald \Big(-\frac{31 \NGen}{40}-\frac{87 \z3}{100}-\frac{5479}{9600}\Big)
+\alb^2 \alu \Big(-\frac{63 \NGen}{8}-\frac{189 \z3}{4}+\frac{7299}{128}\Big)
\nonumber\\
&
+\alb^2 \ald \Big(-\frac{63 \NGen}{8}-\frac{189 \z3}{4}+\frac{7299}{128}\Big)
+\alb^2 \ale \Big(-\frac{21 \NGen}{8}-\frac{63 \z3}{4}+\frac{2433}{128}\Big)
\nonumber\\
&
+\ala \alb \ale \Big(\frac{54 \z3}{5}-\frac{1233}{320}\Big)
+\ala \alb \ald \Big(\frac{2013}{320}-\frac{27 \z3}{5}\Big)
+\ala \ale^2 \Big(-\frac{27 \z3}{5}-\frac{27}{16}\Big)
\nonumber\\
&
+\ala \ald \alu \Big(\frac{24 \z3}{5}-\frac{417}{40}\Big)
+\ala \alb \lam \Big(\frac{117}{40}-\frac{27 \z3}{5}\Big)
+\ala^2 \lam \Big(\frac{351}{400}-\frac{81 \z3}{50}\Big)
\nonumber\\
&
+\ala \ald^2 \Big(\frac{27 \z3}{5}-\frac{1233}{80}\Big)
+\alb^2 \lam \Big(\frac{117}{16}-\frac{27 \z3}{2}\Big)
+\ala \alu^2 \Big(-\frac{9 \z3}{5}-\frac{957}{80}\Big)
\nonumber\\
&
+\alc^2 \alu \Big(-32 \NGen-24 \z3+\frac{910}{3}\Big)
+\alc^2 \ald \Big(-32 \NGen-24 \z3+\frac{910}{3}\Big)
\nonumber\\
&
+\alb \alc \alu \Big(108 \z3-\frac{489}{4}\Big)
+\alb \alc \ald \Big(108 \z3-\frac{489}{4}\Big)
+\ala \alc \ald \Big(12 \z3-\frac{991}{60}\Big)
\nonumber\\
&
+\alb^2 \alc \NGen \Big(\frac{135}{4}-36 \z3\Big)
+\alb \alu^2 \Big(27 \z3-\frac{1161}{16}\Big)
+\alb \ald^2 \Big(27 \z3-\frac{1161}{16}\Big)
\nonumber\\
&
+\alb \ale^2 \Big(9 \z3-\frac{315}{16}\Big)
+\alc \alu^2 \Big(\frac{15}{2}-72 \z3\Big)
+\alc \ald^2 \Big(\frac{15}{2}-72 \z3\Big)
-\frac{27}{10} \ala \ald \ale
\nonumber\\
&
+\alu^3 \Big(9 \z3+\frac{789}{16}\Big)
+\ald^3 \Big(9 \z3+\frac{789}{16}\Big)
+\ale^3 \Big(3 \z3+\frac{71}{16}\Big)
-\frac{387}{8} \alb \ald \alu
\nonumber\\
&
-\frac{27}{10} \ala \ale \alu
-\frac{27}{2} \alb \ale \alu
-\frac{27}{2} \alb \ald \ale
-\frac{3}{2} \ald \ale \alu
+\frac{831 \ald^2 \alu}{16}
+\frac{831 \ald \alu^2}{16}
\nonumber\\
&
-\frac{135 \alu \lam^2}{2}
-\frac{135 \ald \lam^2}{2}
-\frac{45 \ale \lam^2}{2}
+45 \alu^2 \lam
+45 \ald^2 \lam
+45 \alb \lam^2
\nonumber\\
&
+18 \ale^2 \alu
+18 \ale \alu^2
+18 \ald^2 \ale
+18 \ald \ale^2
+15 \ale^2 \lam
+9 \ala \lam^2
-36 \lam^3\,.
\label{eq:gamH_3}
\end{align}
}

It should be noted that the expressions for $\beta_\lambda$ and $\beta_{m^2}$ are free from gauge-fixing parameters $\xiG,\xiW$ and $\xiB$
	which are present in the renormalization constants for the considered Green functions.
The one- and two-loop corrections are in a full agreement with Refs.~\cite{Machacek:1983fi,Arason:1991ic,Luo:2002ey,Mihaila:2012pz}.
The contributions~\eqref{eq:betalam_3} and~\eqref{eq:betam2_3} coincide with the result of Refs.~\cite{Chetyrkin:2012rz,Chetyrkin:2013wya}.
In the limit of vanishing coupling constants $\ala,\alb,\ald$, and $\ale$, the result for the Higgs field anomalous dimension coincides with the expression presented in Ref.~\cite{Chetyrkin:2012rz}.

To conclude, in this paper we present the expressions for three-loop renormalization group quantities
of the SM Higgs sector, i.e., $\beta_\lambda$, $\beta_{m^2}$. Moreover, we provide the result for the anomalous dimension for the Higgs field.
The former can be used  in a study of high energy behaivoir of the SM parameters.
The latter may be exploited in a more accurate analysis
of the Higgs effective potential (see, e.g., Ref.~\cite{Ford:1992mv,Ford:1992pn}).

\subsection*{Acknowledgments}
The authors would like to thank M.~Kalmykov for stimulating discussions.
This work is partially supported by RFBR grants 11-02-01177-a, 12-02-00412-a, RSGSS-4801.2012.2, JINR Grant~No.~13-302-03.

\end{document}